\documentclass[fleqn,10pt]{wlscirep}
\usepackage[utf8]{inputenc}
\usepackage[T1]{fontenc}
\title{Information-theoretic analysis of multivariate single - cell signaling responses using SLEMI}

\author[1]{Tomasz Jetka}
\author[1]{ Tomasz Winarski}
\author[1]{Karol Niena\l towski}
\author[1]{S\l awomir B\l o\'nski}
\author[1,*]{\\ Micha\l \ Komorowski}
\affil[1]{Institute of Fundamental Technological Research, Polish Academy of Sciences, Warsaw, Poland}

\affil[*]{m.komorowski@sysbiosig.org}


\begin{abstract}
Mathematical methods of information theory constitute essential tools to describe how stimuli are encoded in activities of signaling effectors.  Exploring the information-theoretic perspective, however,  remains conceptually, experimentally and computationally challenging. Specifically, existing computational tools enable efficient analysis of relatively simple systems, usually with one input and output only. Moreover, their robust and readily applicable implementations are missing. Here, we propose a novel algorithm to analyze signaling data within the framework of information theory. Our approach enables robust as well as statistically and computationally efficient analysis of signaling systems with high-dimensional outputs and a large number of input values.  Analysis of the NF-$\kappa$B single - cell signaling responses to TNF-$\alpha$ uniquely reveals that the NF-$\kappa$B signaling dynamics improves discrimination of high concentrations of TNF-$\alpha$ with a modest impact on discrimination of low concentrations. Our readily applicable R-package, SLEMI - statistical learning based estimation of mutual information, allows the approach to be used by computational biologists with only elementary knowledge of information theory. \\

\textbf{Availability:}\\ (i) {\it Supplemental Information} - contains theoretical and experimental methods\\
{\color{white} (i) } \href{https://github.com/sysbiosig/SLEMI/blob/master/paper/SI.pdf}{https://github.com/sysbiosig/SLEMI/blob/master/paper/SI.pdf}\\
(ii) R package SLEMI\\
{\color{white} (ii) }{\hangindent 10em} \href{http://github.com/sysbiosig/SLEMI}{http://github.com/sysbiosig/SLEMI}\\
(iii) {\it User Manual} - contains the documentation of the package\\
{\color{white} (iii) }\href{https://github.com/sysbiosig/SLEMI/blob/master/paper/Manual.pdf}{https://github.com/sysbiosig/SLEMI/blob/master/paper/Manual.pdf}\\
(iv) {\it Testing procedures} - contain step-by-step instructions to assist with the package's installation and running essential functions\\ 
{\color{white} (iv) }\href{https://github.com/sysbiosig/SLEMI/blob/master/paper/TestingProcedures.pdf}{https://github.com/sysbiosig/SLEMI/blob/master/paper/TestingProcedures.pdf}\\

\end{abstract}
\begin{document}

\flushbottom
\maketitle

\thispagestyle{empty}

\section{Introduction}
Biochemical descriptions of cellular signaling require quantitative support to explain how complex stimuli (inputs) are translated and encoded in distinct activities of pathway's effectors (outputs) \cite{purvis2013encoding,antebi2017operational}. Information theory and probabilistic modeling offer an attractive approach
\cite{Brennan:2012cj, waltermann2011information,tkavcik2011information}. Regardless of specific details of a signaling pathway, within information theory, a signaling system can be considered as an input-output device that measures an input signal, $x$, by eliciting a stochastic output, $Y$.    In a typical example, the input, $x$, is the concentration of a ligand that activates a receptor, and the output, $Y$, is the activity of a signaling effector, which might be the nuclear concentration of an activated transcription factor quantified over time \cite{BowsherE1320,levchenko2014cellular,selimkhanov2014accurate}.  As signaling systems are inherently stochastic, the input-output relationship is usually represented by the probability distribution $P(Y|X=x)$.  The overall fidelity of signaling systems is within information theory summarised by the information capacity, $C^*$. The information capacity is expressed in bits, and generally speaking, $2^{C^*}$ represents the maximal number of different inputs that a system can effectively resolve (e.g., different ligand concentrations) \cite{cover2012elements,shannon1948mathematical}. The interest in the unique perspective of information theory is increasing with broader availability of single-cell data \cite{lee2014fold, zhang2017nf,  filippi2016robustness,suderman2017fundamental}.  However, exploring the  approach experimentally remains conceptually and technically challenging \cite{Brennan:2012cj, tkavcik2011information, waltermann2011information }. Moreover, existing theoretical tools are computationally and statistically inefficient to provide a further information-theoretic insight for systems with multiple inputs and outputs.  Also, we lack readily applicable implementations. Here, we propose a novel algorithm, that is computationally efficient, provides accurate estimates for relatively small sample size, and, hence, can provide novel biological insight for systems with highly-dimensional outputs and a large number of input values. We also provide the algorithm's robust implementation.

In a typical experiment aimed to quantify how much information can flow through a given signaling system, input values  $x_1\leq x_2...\leq x_m$, ranging from 0 to saturation are considered \cite{Cheong:2011jp, zhang2017nf,selimkhanov2014accurate}. Then, responses to each input level, $x_i$, are quantified in a large number of individual cells. Responses of individual cells are represented as vectors $y_j^i$, where $j$ varies from 1 to the number of quantified cells, denoted as $n_i$. Often, vectors $y_j^i$ contain entries that quantify activities of signaling effectors over time.  The data are assumed to follow an unknown distribution, $y_j^i \sim P(Y|X=x_i)$, which represents the system's input-output relationship. To estimate the information capacity, existing algorithms 
\cite{blahut1972computation, arimoto1972algorithm,vontobel2003generalized, tkavcik2008information, Cheong:2011jp,selimkhanov2014accurate} utilise the data, $y_j^i,$ to construct approximations, $\hat{P}(Y|X=x_i)$, of the output distributions, $P(Y|X=x_i)$, for $i$ ranging from $1$ to the number of input values $m$. 
Thereafter, the approximations, $\hat{P}(Y|X=x_i)$, are  used to evaluate the mutual information (MI)
\begin{eqnarray}\label{eq:mi3}
MI(X,Y) \approx\sum_{i=1}^m \int_{\mathbb{R}^k} \hat{P}(y|X=x_i) P(x_i)  \log_2\frac{\hat{P}(y|X=x_i)}{\hat{P}(y)}dy,
\end{eqnarray}
where $P(x_i)$  is a distribution of input values, which is usually set depending on the context, and $\hat{P}(y)=\sum_{i=1}^m \hat{P}(y|X=x_i) P(x_i) $ is the approximation to the marginal distribution of the output, $Y$.  The maximization of MI with respect to usually unknown probabilities of input values $P(x_i)$ allows computation of the information capacity defined as
\begin{equation}\label{eq:capacity}
C^*=\max_{P(x_1),...,P(x_m)} MI(X,Y).
\end{equation}
The available algorithms differ in the way, in which, the approximation of the output distributions, $\hat{P}(y|X=x_j)$, is constructed. Specifically, Blahut - Arimoto (BA) algorithm \cite{vontobel2003generalized, blahut1972computation, arimoto1972algorithm,Cheong:2011jp} uses a discrete approximation.  All possible values of responses are divided into a finite set of intervals and frequencies of responses falling into the same interval as $y_j^i$ are used as the approximation of ${P}(y_j^i|X=x_j)$. On the other hand, methods based on the small noise approximation assume Gaussian output with a limited variance \cite{tkavcik2008information, dubuis2013positional, tkavcik2014positional, crisanti2018statistics}. Finally, the approach of  \cite{selimkhanov2014accurate}, following earlier work \cite{kraskov2004estimating}, uses the k-nearest neighbors (KNN) method, in which continuous approximations ${P}(y_j^i|x_i)$ are constructed based on the distance of $y_j^i$ to the $k$-th most similar response. Each of the above approaches is practically limited by the dimensionality of the output, Y. The BA algorithm can be essentially applied to systems with the one-dimensional output.  On the other hand, for multidimensional outputs, an accurate estimation of $P(Y|X=x_i)$ using KNN requires a large sample size \cite{mack1979multivariate}.  Moreover, KNN demands arbitrary specification of the parameter $k$, which for insufficient data size does not generally guarantee unbiased estimation \cite{wang2009divergence, kraskov2004estimating,mack1979multivariate,kinney2014equitability}, and yields estimation sensitive to arbitrary assumptions. Moreover, KNN based approaches often require solving computationally expensive optimisation problems.  In Section 1 of the {\it Supplemental Information} ({\it SI}), we provide more background on information theory and existing computational tools. Here further,  we introduce an alternative framework that allows efficient, in terms of sample size and computational time, estimation of the information capacity for systems with high dimensional outputs, $Y$. In addition, our approach uniquely provides probabilities of correct discriminations between different input values. The framework reveals that NF-$\kappa$B signaling dynamics improves discrimination of high concentrations of TNF-$\alpha$ with a modest impact on discrimination of low concentrations. A robust implementation of the proposed computational tools is also provided.

\section{Results}
\subsection{Efficient computation of the information capacity}
In contrast to existing approaches, instead of estimating highly dimensional conditional output distributions $P(Y|X=x_i)$, we propose to estimate the discrete,  conditional input distribution, $P(x_i |Y=y)$, which is known to be a simpler problem \cite{friedman2001elements, silverman2018density}. Estimator of  $P(x_i |Y=y)$, denoted here further as 
$\hat{P}(x_i |Y=y)$, can be built by using bayesian statistical learning methods, here specifically, logistic regression.  Estimation of the MI using $\hat{P}(x_i|Y=y)$ rather than $\hat{P}(y |X=x_i )$ is possible as the MI (Eq. \ref{eq:mi3}), can be alternatively written as
 \cite{cover2012elements}
\begin{eqnarray}\label{eq:mi_inverese}
MI(X,Y) \approx  \sum_{i=1}^m P(x_i) \sum_{j=1}^{n_j} P(y_j^i|X=x_i)  \log_2\frac{\hat{P}(x_i|Y=y_j^i)}{P(x_i)}.
 \end{eqnarray}
Therefore, for a given ${P}(x_i)$ and $\hat{P}(x_i|Y=y)$ MI can be evaluated without knowledge of $P(Y|X=x_i)$. Although $P(Y|X=x_i)$ is still present in the above sum, it represents averaging of  the term $ \log_2\frac{\hat{P}(x_i|Y=y_j^i)}{P(x_i)}$ over available data, which can be achieved with data $y_j^i$ alone, without explicit knowledge of
$P(y|X=x_i)$.  Further, the above formulation allows to employ an efficient convex optimization scheme to compute $C^*$ from MI. Therefore, no numerical gradient optimization is needed.  In Section 2 of the {\it SI} we describe the algorithm in detail and prove its mathematical correctness. In Box S1 we present the algorithm as pseudocode. 

The logistic regression used to approximate $\hat{P}(x_i |Y=y )$ combined with the convex optimization led to the algorithm that outcompetes existing approaches in terms of sample size needed to provide accurate estimates, computational time, and robustness to algorithm settings.   As opposed to the KNN based approaches, algorithm's settings do not impact the estimates, which ensures robust estimation. The approach is also an order of magnitude faster compared to KNN method, especially for systems with high number of input values, $m$. These advantages are demonstrated in Section 3 of the {\it SI}, specifically in Figures S1 and S2. The benefits are of particular importance for signaling systems with multidimensional outputs, $Y$, and a large number of considered input values, $m$.

Importantly, the algorithm, also, uniquely allows analyzing signaling systems in terms of discrimination error. Precisely, the information capacity {\it per se} does not tell us, which input values cells can effectively distinguish. It only provides an overall measure of signaling fidelity. Given that our approach is based on the approximation of the conditional input distribution, $\hat{P}(x_i|y)$, the probabilities of correct discrimination are readily available. It can be shown \cite{cover2012elements} that the strategy that maximizes the probability of correct guessing, which input $x_i$ lead to observed output $y$, is the maximum \textsl{a posteriori} rule, which selects $x_i$ with  highest  $\hat{P}(x_i|Y=y)$.  We describe the calculation of the probabilities of correct discrimination in Section 4.5 of the {\it SI}.

The advantages of the proposed framework extend beyond statistical and computational aspects. To demonstrate this, we have experimentally measured, single - cell signaling responses of the 
 NF-$\kappa$B system to a range of concentrations of TNF-$\alpha$ in murine embryonic fibroblasts cell line. Analysis of the experimental data revealed how information transfer is distributed over time. It also uniquely showed that the dynamics of the NF-$\kappa$B signaling responses leads to improved discrimination of high TNF-$\alpha$ concentrations with a minor effect on discrimination of low concentrations. 

\subsection{Signaling dynamics of NF-$\kappa$B system strongly improves discrimination of only high TNF-$\alpha$ concentrations}

\begin{figure}[!h]
\includegraphics[width=0.75\paperwidth]{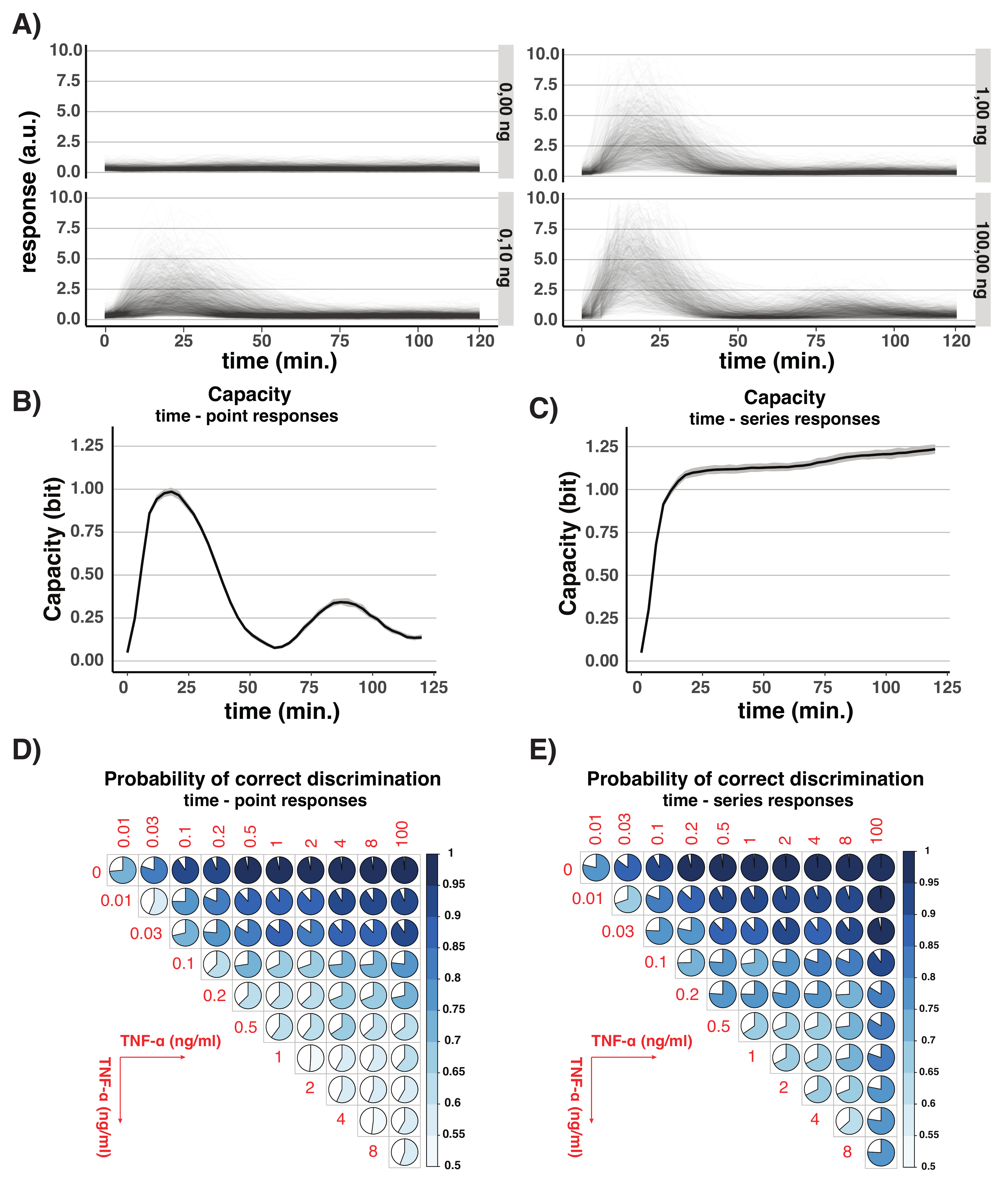}
\caption{
Information-theoretic analysis of the NF-$\kappa$B responses to TNF-$\alpha$ stimulation. {\bf (A)} Temporally resolved responses of individual cells to selected concentrations of TNF-$\alpha$. The panel corresponds to Fig S4. {\bf (B)} Information capacity as a function of time for time-point responses (see text). {\bf (C)} As in (B) but for time-series responses.
{\bf (D)} Probabilities of the correct pairwise discrimination between TNF-$\alpha$ concentrations for time-point responses. The color filled fraction of the circle marks the probability of correct discrimination (see text and {\it SI}). 
{\bf (E)} The same as in (D) but for time-series responses. 
Modeling details:  Uncertainties of estimates (grey ribbons in B and C) were obtained by bootstrapping 80\% of data (repeated 100 times). Probabilities in D and E present mean of 50 bootstrap re-sampling. 
} 
\end{figure}

The NF-$\kappa$B signaling is one of the key pathways involved in the control of the immune system
\cite{tay2010single,oeckinghaus2011crosstalk, sun2017non}. It is also one of the first cellular signaling systems studied within the framework of information theory\cite{Cheong:2011jp}. So far, several papers quantified its information capacity, e.g., \cite{selimkhanov2014accurate,Cheong:2011jp, zhang2017nf}. Interestingly, response dynamics have been shown to have greater signaling capacity compared to non-dynamic responses \cite{zhang2017nf, selimkhanov2014accurate}. To demonstrate what novel insight can be gained with our framework, we have measured NF-$\kappa$B responses ($y^i_j$'s in the above notation) to a range of 5 minutes pulses of TNF-$\alpha$ concentrations ($x_i$'s), in single - cells, using life confocal imaging. Experimental methods are described in Sections 4.1 - 4.3 of the {\it SI}. Fig. 1A  shows temporally resolved responses,  $y^i_j$, to representative four concentrations, whereas Fig. S4 to all considered concentrations. In order to provide a further insight into the dynamic aspect of signaling, we used the data to calculate the information capacity for two different scenarios: time-point and time-series responses. For time-point responses, we consider experimental measurements at a specified time only. On the other hand, for time-series responses, we consider measurements from the beginning of the experiment till an indicated time. Fig. 1B and C show information capacity for time-point and time-series responses, respectively, as a function of time. Time-series data include time-point data, which implies higher information content. Precisely, information capacity for time-series responses increases sharply over 1 bit at $\sim$ 25 min., and reaches $\sim$1.3 bits at late times, i.e., $\sim$120 minutes. In contrast, information capacity for time-point responses reaches 1 bit around the time of maximal response, i.e., $\sim$20 min, only, and remains below 1 bit for all other times. Interestingly, time-point responses exhibit a second peak of information transfer at $\sim$85 minutes. This is an extension of the result of  \cite{selimkhanov2014accurate} and \cite{zhang2017nf}, where signaling dynamics, represented by time-series, have been shown to increases the information capacity. The efficiency of the algorithm, uniquely, allowed to calculate the capacity as the function of time, and, hence,  reveal how the information transfer is distributed over time. These computations involved outputs containing up to 40 entries, which is usually not achievable with other approaches. Most importantly, however, with our approach, we can decipher what information is transferred using the additional $\sim0.3$ bits provided by the response dynamics. 
To address this, we have calculated the probabilities of correct discrimination between all concentration pairs. This can be done within our framework as it is based on the estimation of the conditional input distribution, $\hat{P}(x_i|Y=y)$. For each single - cell response $y_j^i$, we found most likely input value and compared whether the true one is the one most likely. When most likely value matched the correct one we interpreted this as the correct discrimination. Calculated probabilities of correct pairwise discrimination for time-point and time-series responses are presented as pie-charts in Fig.1D and Fig.1E, respectively. Random discrimination yields 0.5 chance of correct guessing. Hence, all probabilities are $\geq 0.5$. Comparison of Fig.1D and Fig. 1E,  demonstrates that time-point and time-series responses result with similar probabilities of correct discrimination between low and high concentrations, i.e., pie-charts close to full circle. Also,  probabilities  of correct discrimination between low concentrations are similar in both scenarios. On the other hand, discrimination between high concentrations is largely improved for time-series responses. For instance, discrimination between 0 and 100 ng/ml is close to perfect for both scenarios.  On the other hand, discrimination between 8 and 100 ng/ml based on time-point responses is close to random, whereas it is more than 75\%  successfully based on time-series responses. This demonstrates that signaling dynamics contains information that improves discrimination of high TNF-$\alpha$ concentrations,  which is uniquely revealed by our computational approach. In the light of this analysis, the higher capacity of time-series responses results largely from improved discriminability of high concentrations rather than improved discriminability of all concentrations. Sections 4.3 - 4.5 of the {\it SI} contain more details on the analysis of experimental data.
\clearpage
\subsection{R-package}
Our algorithm is available as robustly implemented R-Package SLEMI. It can be used by a computational biologist with a limited background in information theory. Details on installation and applicability are provided in the package's {\it User Manual}.
Step-by-step {\it Testing Procedures} are also provided to assist with package's installation and running essential functions.  The package includes the  NF-$\kappa$B dataset as well as scripts to reproduce Fig. 1. Computations needed to plot each panel of the figure, without bootstrap, do not exceed several minutes on a regular laptop.

\section{Summary \& conclusions}
Building upon existing approaches, our framework considerably simplifies information-theoretic analysis of multivariate single-cell signaling data.  It benefits from a novel algorithm, which is based on the estimation of the discrete input distribution as opposed to the estimation of continuous output distributions. Conveniently, the algorithm does not involve numerical gradient optimization. These factors result not only in short computational times but, also, in relatively low sample sizes needed to obtain accurate estimates. Therefore, our framework is particularly suitable to study systems with high dimensional outputs and a large number of input values. 
Also, the approach relates the information capacity to the probability of discrimination between different input values. 

The overall molecular and biochemical mechanisms how individual cells transmit signals to effectors are widely understood \cite{nurse2008life}. However, we lack an understanding of how the stimuli are translated into distinct responses and, hence, how to effectively control cellular decisions and processes \cite{nurse2008life, antebi2017operational, suderman2017fundamental}. Specifically, the induction of distinct responses in individual cells by means of biochemical interventions is most often problematic \cite{behar2013dynamics, rue2015cell, symmons2016s}.  Results of our work appear to contribute a relevant tool to apply information-theoretic analysis to more complex data sets on signaling systems than achievable with available approaches. A more insightful information-theoretic perspective is necessary to address the question of how cells transmit information about identity and quantity of stimuli, and further how signaling systems enable cells to perform complex functions. 

\section*{Acknowledgements}
The immortalised murine embryonic fibroblasts cell line (3T3) expressing fluorescent fusion proteins relA-dsRed was kindly provided by prof. Savas Tay. The focus on the NF-$\kappa$B pathway was inspired by prof. Tomasz Lipniacki. Experimental component of this research was carried out with the use of CePT 
infrastructure financed by the European Regional Development Fund within  the Operational Program {\it Innovative Economy} for 2007-2013. TJ was supported by his own funds and the European Commission Research Executive Agency under grant CIG PCIG12-GA-2012- 334298, TW by IUVENTUS PLUS grant IP2012016572, MK by the Polish National Science Centre under grant 2015/17/B/NZ2/03692.

\bibliography{literature}
\end{document}